\documentclass[12pt]{iopart}
\usepackage{iopams}          
\usepackage{graphicx}        
\usepackage{amsmath}
\usepackage{hyperref}
\hypersetup{
    colorlinks=true,
    linkcolor=blue,
    citecolor=blue,
    urlcolor=blue,
    }
\newcommand{\ex}{\mathop{\mathbb{E}}}

\begin{document}

\title[Reconstruction-free magnetic control of DIII-D plasma with deep reinforcement learning]{Reconstruction-free magnetic control of DIII-D plasma with deep reinforcement learning}

\author{G F Subbotin$^1$, D I Sorokin$^1$, M R Nurgaliev$^1$, A A Granovskiy$^1$, I P Kharitonov$^1$, E V Adishchev$^1$, E N Khairutdinov$^1$, R Clark$^2$, H Shen$^3$, W Choi$^3$, J Barr$^3$, D M Orlov$^2$}
\address{$^1$Next Step Fusion, Bertrange, L-8070, Luxembourg}
\address{$^2$Center for Energy Research, University of California San Diego, CA 92093, United States}
\address{$^3$General Atomics, San Diego, CA, 92186, United States}
\ead{gs@nextfusion.org}

\begin{abstract}
Precise control of plasma shape and position is essential for stable tokamak operation and achieving commercial fusion energy. Traditional control methods rely on equilibrium reconstruction and linearized models, limiting adaptability and real-time performance. Here,the first application of deep reinforcement learning (RL) for magnetic plasma control on the mid-size DIII-D tokamak is presented, demonstrating a nonlinear approach that improves robustness and flexibility across plasma scenarios. Using the Soft Actor-Critic algorithm, this method eliminates the need for equilibrium reconstruction, enabling high-speed control execution and scalability on larger fusion devices. NSFsim, a 2D Grad-Shafranov equilibration solver with a circuit equation and a 1D transport solver, is used to train the agent. Its capability of reproducing the kinetic parameter evolution alongside magnetic equilibria evolution appears to be an essential factor significantly affecting control quality. RL-based controllers demonstrated robust magnetic control in experimental application at DIII-D, preserving control performance in transient events during plasma discharges, and reaching target parameters from the first discharge without additional tuning or modifications. The approach itself has significant generalization potential across devices and targets. This work represents a step toward AI-driven, real-time plasma control, advancing the feasibility of next-generation fusion reactors.

\end{abstract}
\submitto{\NF}

\maketitle

\section{Introduction}

Nuclear fusion is a promising energy source that offers the potential to provide clean, safe, carbon-free, and nearly limitless power to meet global energy demands. Several approaches have been developed to achieve the high temperatures required for fusion reactions since the 1950s. One of the most successful machines that is based on the magnetic confinement of hot plasma is the tokamak. In tokamaks, an external magnetic field is created by both toroidal (TF) and poloidal (PF) magnetic field coils. While TF coils, together with the toroidal plasma current, create rotational transform -- helical trajectories of magnetic field lines around the torus -- that play a crucial role in plasma confinement, stabilization, and ensuring quasi-axisymmetric behavior, PF coils create an additional external poloidal magnetic field for active plasma control. PF coils are used for many applications, such as generating plasma currents, plasma shape and position controlling, maintaining stability, assessing divertor configurations, and exhaust control. Achieving high-performance plasma regimes and maintaining stability at the same time require real-time control systems. Since instabilities in tokamak plasma can grow rapidly on a millisecond timescale, the plasma control system (PCS) must be fast yet precise to control plasma efficiently.

Present algorithms utilize equilibrium reconstruction techniques for shape control. Based on magnetic diagnostics, codes such as rtEFIT~\cite{rtEFIT}, LIUQE~\cite{Moret2015}, or Equinox~\cite{Blum2008} are able to obtain the spatial distribution of poloidal flux and toroidal current density by searching for an approximate solution to the Grad-Shafranov equation that fits best diagnostics measurements. These codes achieve a processing speed of 1-2 ms per timestep through parallel programming while balancing computational efficiency and accuracy. A groundbreaking approach was introduced in work \cite{Degrave_2022} utilizing reinforcement learning (RL) algorithms to train a nonlinear controller that directly uses raw magnetic diagnostic signals as inputs to generate control commands. This method simplifies the controller development process and shifts the control paradigm from a traditional ``sensor-actuator'' model to a more advanced ``state-oriented'' perspective.

This paper considers an application of an RL-based approach in the context of magnetic control of high-performance discharges on the DIII-D tokamak \cite{Thome_2024, Holcomb_2024}. The control task is modeled as a partially observable Markov decision process (POMDP), and the Soft Actor-Critic~\cite{SAC} algorithm is applied to train an agent. An asymmetric Actor-Critic architecture with privileged information, which was used in this work, stabilizes and accelerates the training process in the presence of uncertainties in the environment dynamics. NSFsim is used in the training process to accurately reproduce tokamak behavior during a plasma shot. By simultaneously simulating both the magnetic and kinetic aspects of plasma equilibrium evolution, the simulator provides a realistic plasma response to the agent. This enables tracking of plasma dynamics during pressure decay in the absence of auxiliary heating while allowing the agent to learn how to robustly maintain plasma across various heating conditions. This effect was clearly observed in the experimental data analysis and is discussed in the final section of this paper. Trained controllers are successfully tested in various plasma conditions, including the high-performance H-mode regime \cite{Wagner_2007} and in transient events such as H-L transitions \cite{Bourdelle_2015}. Results show high robustness of the controllers in out-of-domain tests with varying auxiliary heating power.

This paper is organized as follows. Section \ref{sec:mag_control} formulates the RL task for plasma shape and position control, addressing both machine learning and physics aspects of the problem. Section \ref{sec:implementation} considers integrating the controlling agent into the existing DIII-D plasma control system. Experimentally validating the approach and presenting the data analysis in section \ref{sec:results}, which is followed by a discussion in section \ref{sec:discussion}. 

\section{Magnetic plasma control as a reinforcement learning task} \label{sec:mag_control}

The standard approach to magnetic control employs real-time equilibrium reconstruction techniques that use magnetic probe, flux loop, coil current, and plasma current measurements to obtain the 2D distribution of the poloidal flux function $\psi(R,Z)$ and, consequently, a plasma boundary $(R_\mathrm{b}, Z_\mathrm{b})$. This allows for the estimation of control variable errors between the current plasma shape and the given targets that can be specified by selected control points. Calculated errors are fed to PID-based controllers that produce control commands to actuators. The main limitation of this approach is the reliance on additional shape reconstruction algorithms and the need to tune the PID gains for different plasma shapes.

We view magnetic plasma control from a reinforcement learning perspective and propose a single RL-controller that directly processes raw values of magnetic probes, magnetic loops, and coil currents and outputs actuator commands. Interaction between an agent (RL-controller parametrized by neural network) and environment (simulated or physical tokamak) is considered as POMDP since the environment dynamics also depend on the kinetic parameters of the plasma, which are rarely accessible in real-time. We train the agent in a simulated environment to learn a policy $\pi(a|o^\mathrm{rt})$ which maps the noisy values of real-time sensors (``$o^\mathrm{rt}$’’ -- real-time observations) to actuator commands (``$a$’’ -- actions).

\begin{figure}[ht]
    \centering
    \includegraphics[width=1.0\linewidth]{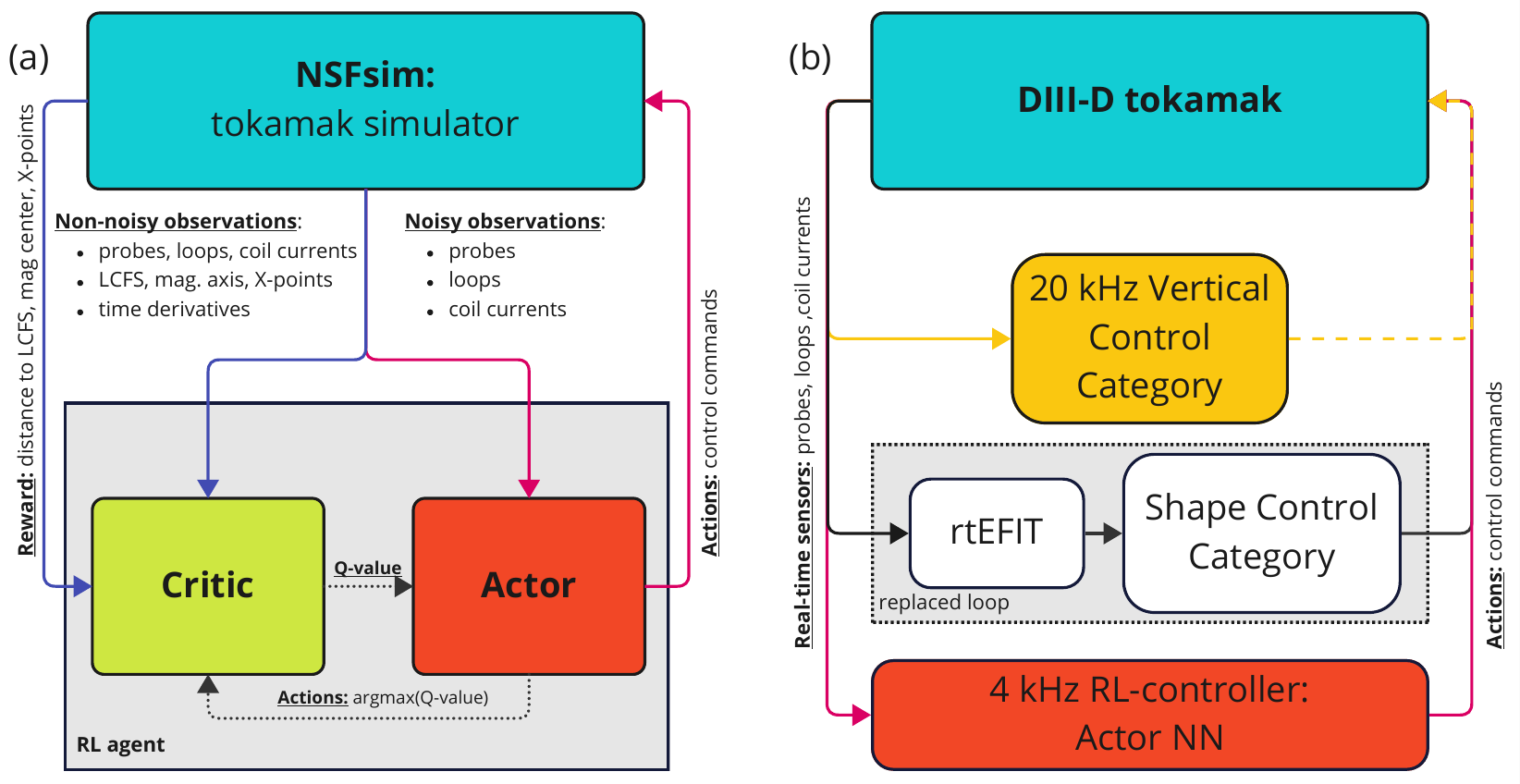}
    \caption{(\textbf{a) The training environment}: NSFsim, a 2D Grad-Shafranov and transport solver, provides responses to the agent's actions. The agent comprises two neural networks -- the Critic and the Actor, which observe the magnetic state asymmetrically. The Actor observes raw, noisy signals from magnetic probes, flux loops, and coil currents, while the Critic has access to non-noisy signals and additional information, including the time derivatives and coordinates of the LCFS, magnetic axis, and X-points. The Critic predicts the reward, which is constructed as a combination of shape and position distances, and the Actor seeks to maximize the reward.
    \textbf{(b) Test environment}: The Actor neural network operates in a 4 kHz feedback loop, replacing the standard shape control loop, including the rtEFIT equilibrium reconstruction. Additional fast vertical control can be switched off during the experiment, as illustrated by the dashed arrow. The Actor generates low-level control commands for the choppers. Elements with the same color correspond between figures (a) and (b).}
    \label{fig:train_test}
\end{figure}

The principal scheme of the training process is given in figure~\ref{fig:train_test}(a). Plasma evolution is modeled using the NSFsim~\cite{nsfsim} simulator, which solves the free-boundary Grad-Shafranov equation coupled with 1D core transport. NSFsim provides plasma response by PF-coil current changes. The RL-agent is trained through interaction with such an environment by the Soft Actor-Critic algorithm \cite{SAC} with asymmetric Actor-Critic architecture. The Actor network observes noisy values of magnetic probes, loops, and coil currents, while the Critic network has access to privileged information, e.g., exact values of magnetic probes, magnetic loops, coil currents, last closed-flux surface (LCFS) shape, magnetic center, X-points, and their time derivatives. The Actor produces low-level control commands that nonlinearly establish current values in shaping PF-coils. More detailed information on Actor commands is given in the next section.

During training, the agent learns to maximize the cumulative discounted reward. The reward reflects proximity between the current and target plasma shapes, i.e., it equals $1$ if the plasma shape matches the target shape and decreases to $0$ when the shape becomes significantly different. The Critic $Q^{\pi}(o^\mathrm{pri}, a)$ is parametrized by a neural network of size 552 $\times$ 256 $\times$ 1 with a ReLU activation function. It predicts the cumulative discounted reward plus the entropy term. The Actor learns to maximize the cumulative reward predicted by the Critic network. It is designed to be a compact and fast neural network that consists of a 3-layer MLP of size 132 $\times$ 256 $\times$ 18 and a ReLU activation function. The Actor's policy  $\pi(a|o^\mathrm{rt})$ provides a direct mapping from real-time sensors to actuator commands. Therefore, a successfully trained Actor can be deployed directly to the DIII-D PCS and run without any need to calculate an equilibrium reconstruction. 

Each simulation episode begins with an initial plasma state that includes the following parameters: plasma current, toroidal magnetic field, PF-coil currents, gradient profiles ($dp/d\psi$ and $FdF/d\psi$, where $p$ is plasma pressure and $F$ is toroidal flux function), electron ($T_\mathrm{e}$) and ion ($T_\mathrm{i}$) temperatures, and effective ionic charge ($Z_{\mathrm{eff}}$). We extracted this information from experimental data and EFIT \cite{Lao_2005} reconstructions for a selected set of shots. We selected a specific timestamp in one of the selected shots to define a reference plasma shape that should be maintained by the agent. Due to measurement errors and possible inconsistencies of different diagnostic data, the initial state reconstructed by NSFsim has a slightly different $\psi(R,Z)$ distribution than, for example, was obtained via EFIT or GSevolve~\cite{Welander_2019} simulations. This difference is studied in work~\cite{nsfsim}.

To facilitate robust sim-to-real transfer, the agent should encounter numerous variations of plasma and hardware conditions. This variation, in fact, defines a domain of robust applicability of the agent. An out-of-domain application is still possible, but it will be characterized by reduced precision of control, as it will be shown in section \ref{sec:results}. In this work, two different ways of plasma and hardware conditions treatment were applied. First, each training episode starts with a disturbed initial plasma state. Values of $T_\mathrm{e}$, $T_\mathrm{i}$, and $Z_\mathrm{eff}$ at the magnetic axis, $\rho=0$, and the separatrix, $\rho=1.0$, are varied according to the min-max intervals given in Tab.~\ref{tab:random_episode}. Additionally, variation of initial coil currents in the range of $\pm100$~ A allows us to assess different initial plasma shapes from which the RL-controller will try to guide plasma to a given target. After initialization, magnetic and kinetic plasma evolution is modeled with NSFsim. Two strategies for the simplified treatment of heat and particle transport are performed. In the first one, the transport coefficients are given by the Bohm-gyroBohm model~\cite{Erba_1998}, and plasma is considered to be Ohmically heated without specifying auxiliary power and current drive deposition. Instead, randomized initial electron and ion temperatures are varied to imitate both hot and cold plasmas. Such treatment of transport allows the RL-agent to be robust to different heating conditions and save computational time since no additional heating and current drive calculations are involved. Further, at each time step, Gaussian noise is added to the observed values of probes, loops, and coil currents with zero mean and standard deviation listed in Tab.~\ref{tab:random_step}. These standard deviation values were obtained through the analysis of measured signals. In such conditions, Agent observed decreasing temperature during each training episode, providing operations in the context of different $\beta$ values. Agents trained with this procedure will be marked as RL24 later in this article.
The second strategy implies the use of fitted measured profiles of electron temperature, electron density, and ion temperature. These profiles are scaled by a fixed multiplier for each training episode. Multipliers were varied in range (0,  4]. Keeping kinetic profiles constant and, consequently, constant $\beta$ values throughout each training episode allows better handling of the Shafranov shift, as it will be shown later. Agents trained with this procedure will be marked as RL25 later in this article.
A set of four RL controllers was prepared for testing in this way:
\begin{itemize}
  \item RL24\#1 was trained using H-mode reference shot \#200563 with 56 probes and 38 loops
  \item RL24\#2 was trained using L-mode reference shot \#186093 with 56 probes and 38 loops
  \item RL25\#1 was trained with H-mode reference shot \#193593 with 48 probes and 30 loops
  \item RL25\#2 was trained with H-mode reference shot \#193593 with 46 probes and  23 loops
\end{itemize}

The described methods allow training a single controller in less than 24 hours on a small cluster with 512 CPU threads. The developed pipeline is generally machine agnostic and requires only a pre-configured tokamak simulator with an actuator layer model.

\begin{table}[ht]
\centering
\begin{tabular}{|c|c|c|c|c|c|c|}
\hline
 & $T_e(0)$, eV & $T_e(1.0)$, eV & $T_i(0)$, eV & $T_i(1.0)$, eV & $z_{\mathrm{eff}}(0)$ & $z_{\mathrm{eff}}(1.0)$\\
\hline
min & 1000 & 10 & 1000 & 10 & 1 & 1 \\
max & 5000 & 300 & 5000 & 300 & 4 & 4\\
\hline
\end{tabular}
\caption{\label{tab:random_episode} Dynamics randomization parameters.}
\end{table}

\begin{table}[ht]
\centering
\begin{tabular}{|c|c|c|c|}
\hline
 & probes, mT & loops, mWb & coil currents, A \\
\hline
std & 0.01 & 0.01 & 100 \\
\hline
\end{tabular}
\caption{\label{tab:random_step}Observation randomization parameters.}
\end{table}

\section{Implementation of the RL Agent in the Plasma Control System at DIII-D} \label{sec:implementation}

RL-controller replaces the standard shape control category. DIII-D utilizes~\cite{Walker_2003} the isoflux~\cite{Eldon_2020} algorithm for magnetic control by specifying a set of shape control target points, including X-points, distributed across the LCFS. Shape control uses a 4~kHz feedback loop and relies on the rtEFIT \cite{rtEFIT} algorithm to reconstruct the location of the LCFS and X-points to calculate control errors. Then, it applies a separate PID regulator to compute actuator commands for 18 shaping PF coils. The vertical position controller uses a 20~kHz control loop and a linear combination of magnetic probes to estimate the position of the magnetic axis, followed by a PID regulator to convert it to actuator commands for the vertical control coils. Commands from both controllers are summed up and passed to the power system. These two loops are illustrated in figure~\ref{fig:train_test}(b). The incorporation of the RL-based controller into the PCS is designed to replace rtEFIT, isoflux control, and fast vertical stability control. However, for operational convenience, the 20 kHz feedback loop can remain active alongside the RL-controller if needed.

To train an end-to-end controller capable of controlling the physical tokamak, we include a model of the actuator system of DIII-D in our simulation environment. The RL-agent needs to interact with PF-coils through control of their choppers, devices that set voltage applied to the coils regarding control voltage input. This conversion is non-linear and requires an additional layer of modeling. The model of the choppers incorporated in GSevolve is implemented in our simulation environment. It provides an average voltage response depending on the command from the RL-agent, coil current, and power supply voltage. These tools provided consistency across domains and allowed the agent to operate almost seamlessly from computation to real-time control.

One additional challenge in this task is handling the flexibility of the DIII-D patch panel and VFI Bus. The VFI bus connects to the backend of a set number of coils according to the patch panel but does not connect to the ground, requiring return coils to accept all excess current; this configuration results in extra complexity from charge getting stored on the VFI bus, impacting the ability for the chopper and power supplies to push more current towards the VFI bus (the reverse is true for negative PF coil currents). Fortunately, this can be modeled once, but the patch panel can vary from shot to shot (usually if a change of shape is needed). For simplicity, a single RL-controller is restricted to be patch panel-specific, which is usually consistent throughout the entirety of the experimental day.

The PCS module is designed in a way that the RL-controller can overwrite the standard control algorithm during the middle of the discharge. It allows us to avoid affecting other scientific programs while testing our controller on the tail end of a shot. When the RL-controller is turned on, it collects the most recent measurement from magnetic diagnostics and outputs a command in the range of [-10, +10], modulating the chopping frequency, allowing the corresponding amount of voltage range through [0\%, 100\%] of power supply voltage; the controller performs this cycle every $250~\mathrm{\mu s}$.

\section{Experimental test results}\label{sec:results}

\begin{figure}
    \centering
    \includegraphics[width=0.5\linewidth]{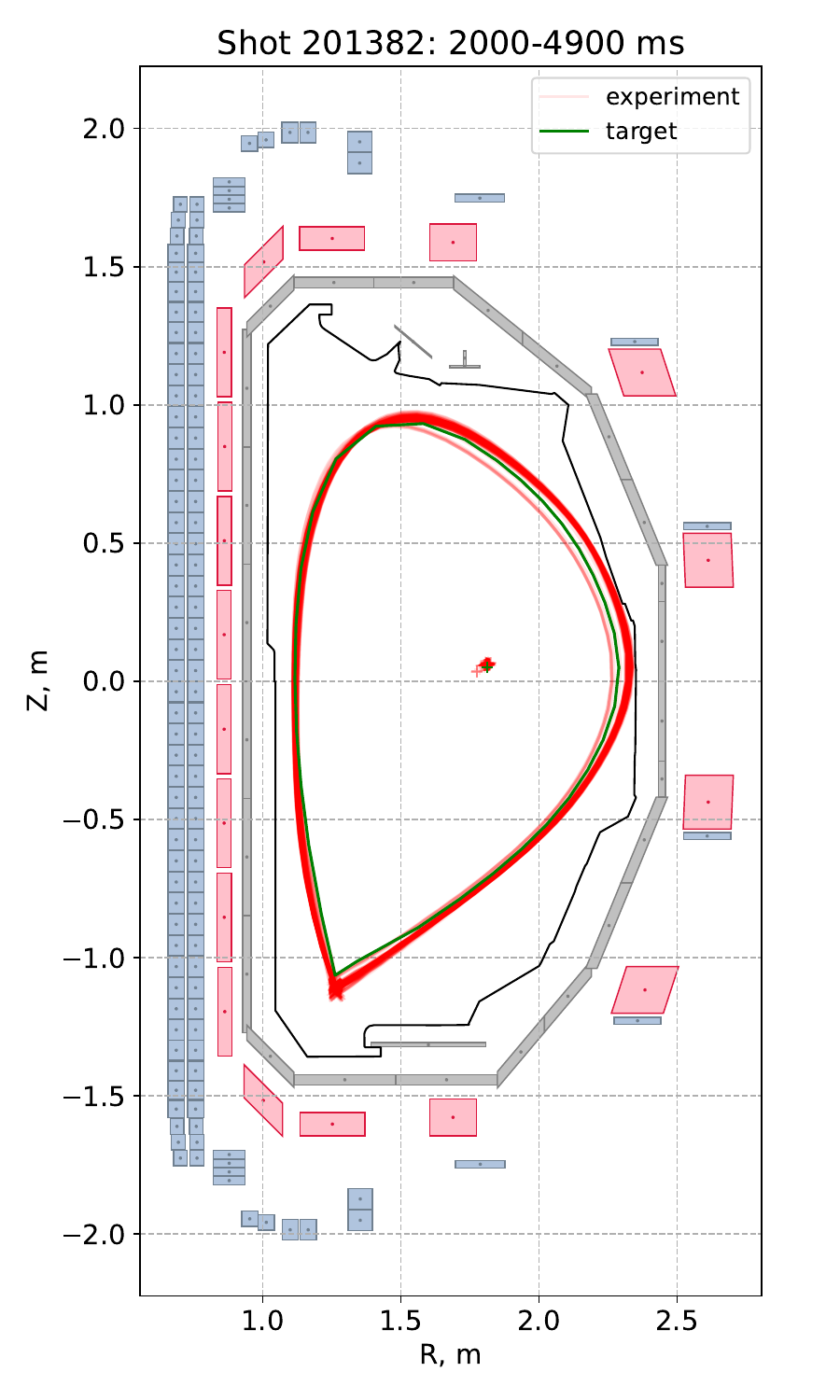}
    \caption{Plasma boundary shape evolution. Red contours show the evolution of the plasma boundary with a 20 ms time step. While each contour is drawn with transparency, their overlap shows a small dispersion around quasi-steady-state equilibrium. The target is given by a green line.}
    \label{fig:201382_dynamic}
\end{figure}

Two series of experiments were carried out. The first session was with RL24 agents. As mentioned above, the patch panel of DIII-D determines which coils are active coils and their coil current direction, which, therefore, forces the RL-controller to be dependent on chopper configuration. Overall, four RL-controllers were trained and tested on three different patch panels. A
First, a general proof-of-concept test was conducted during shot \#201381 using the fast vertical stability algorithm to help sustain the plasma position. Subsequently, in shot \#201382, the control interval was extended from 0.5 to 3 seconds to evaluate the controller's ability to manage plasma longer than during training. In this shot, the 20 kHz vertical stabilization from isoflux was turned off at $t=3$~s. The time series of the plasma boundary for this shot is shown in figure~\ref{fig:201382_dynamic}(a).

\begin{figure}
    \centering
    \includegraphics[width=0.8\linewidth]{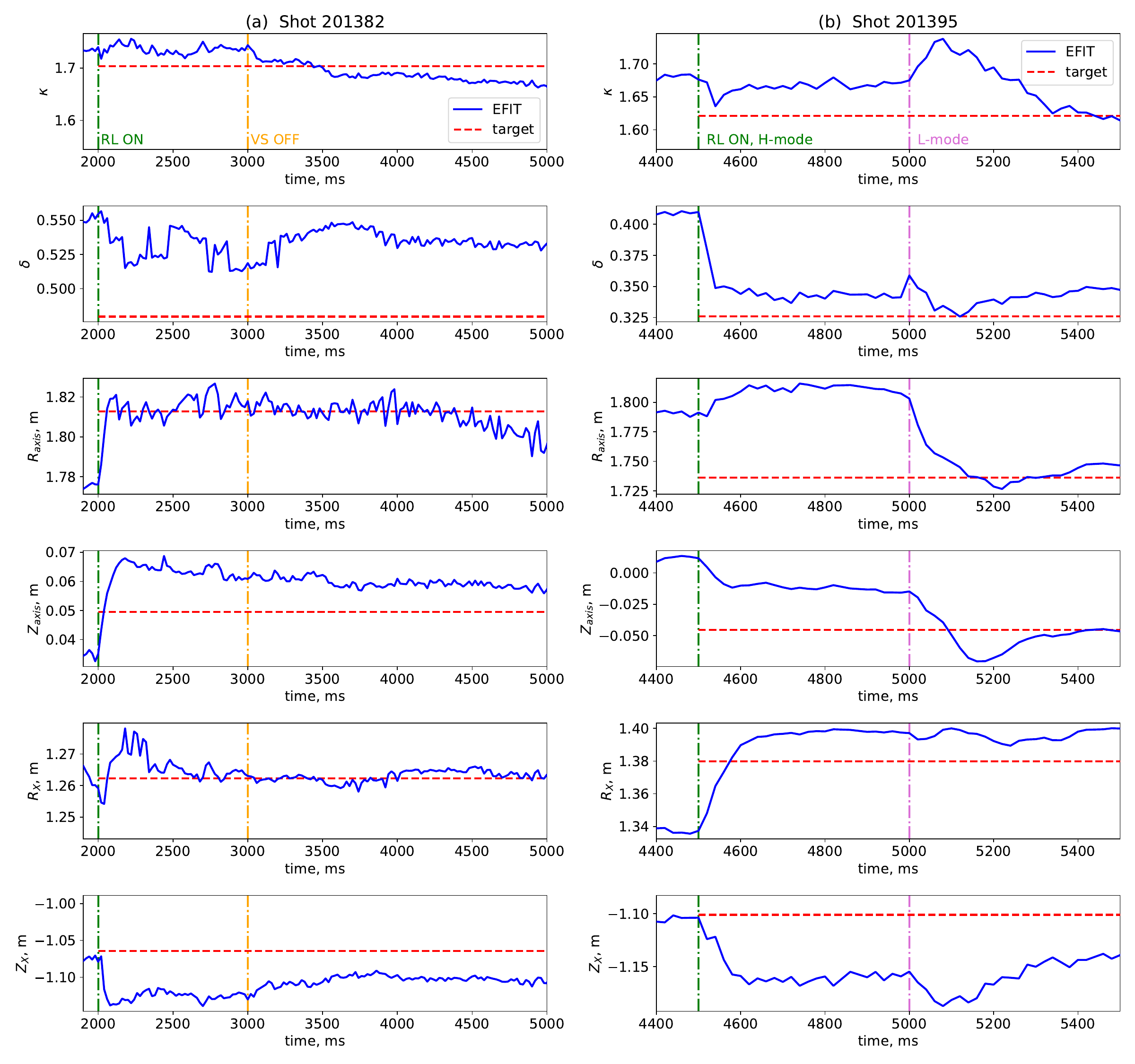}
    \caption{Comparison between the target values that were used for training the RL-controller and the experimental result. Illustrated values are elongation ($\kappa$), up-down average triangularity ($\delta$), coordinates for magnetic axis ($R_{\mathrm{axis}},Z_{\mathrm{axis}}$), and X-point ($R_{\mathrm{X}},Z_{\mathrm{X}}$). (a) Shot 201382 with H-mode regime and 3~s control interval by controller RL24 \#1. At time $t=3$~s, fast vertical stability is turned off. (b) Shot 201395 with H-L transition at $t=5$~s due to switching NBI off.}
    \label{fig:control_quality}
\end{figure}

The RL-controller successfully maintained the plasma center at the target location despite the higher heating power compared to the reference case. However, the outer boundary moved significantly closer to the first wall. This boundary shift resulted from the combination of higher plasma pressure and the RL controller's generation of 15\% lower total PF-coil currents, leading to an increase in plasma volume. Time traces of the shape parameters, as well as the magnetic center and X-point coordinates, are presented in figure~\ref{fig:control_quality}(a). Notably, turning off the 20 kHz vertical stabilization at $t=3$~s did not affect the control quality, which remained consistent until the end of the discharge. The most prominent discrepancy was observed in the $Z$-coordinate of the X-point. 

While the RL24 \#1 was trained on an H-mode reference, \#2 was trained on an L-mode reference with only short NBI blips of 2 MW power. Despite that, the RL24 \#2 managed to control shape in the H-mode plasma with continuous $P_\mathrm{NBI} = 7.5$~MW. Control quality can be seen in figure~\ref{fig:control_quality}(b). Working in a significantly different transport regime results in higher deviations from targets that are eliminated after switching the NBI off.

\begin{figure}
    \centering
    \includegraphics[width=1\linewidth]{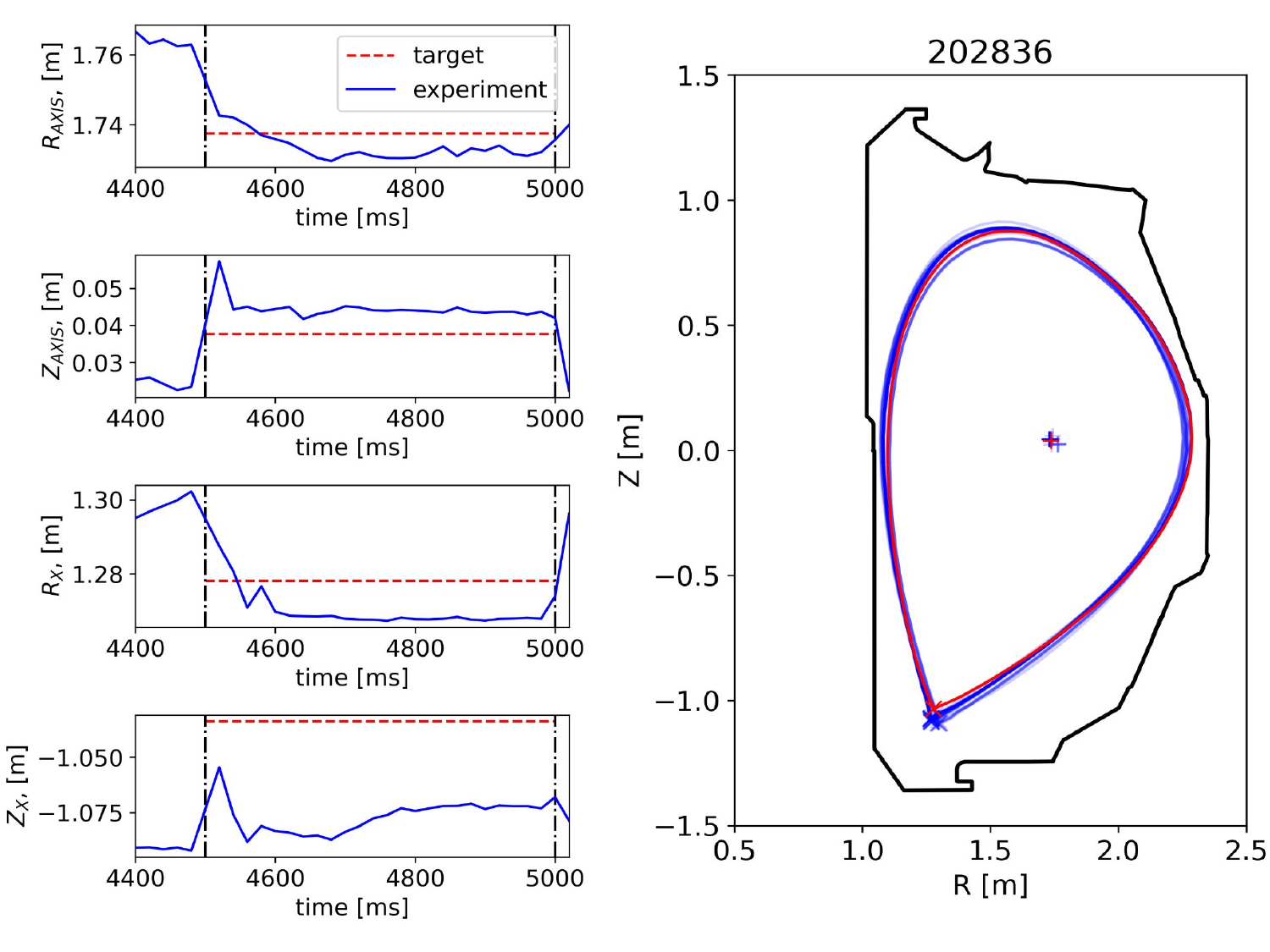}
    \caption{Plasma axis and x-point position timetraces (left) compared with target values (red line) and shape time evolution (right) in blue line compared with target shape (red line)}
    \label{fig:RL25performance}
\end{figure}

\begin{figure}
    \centering
    \includegraphics[width=1\linewidth]{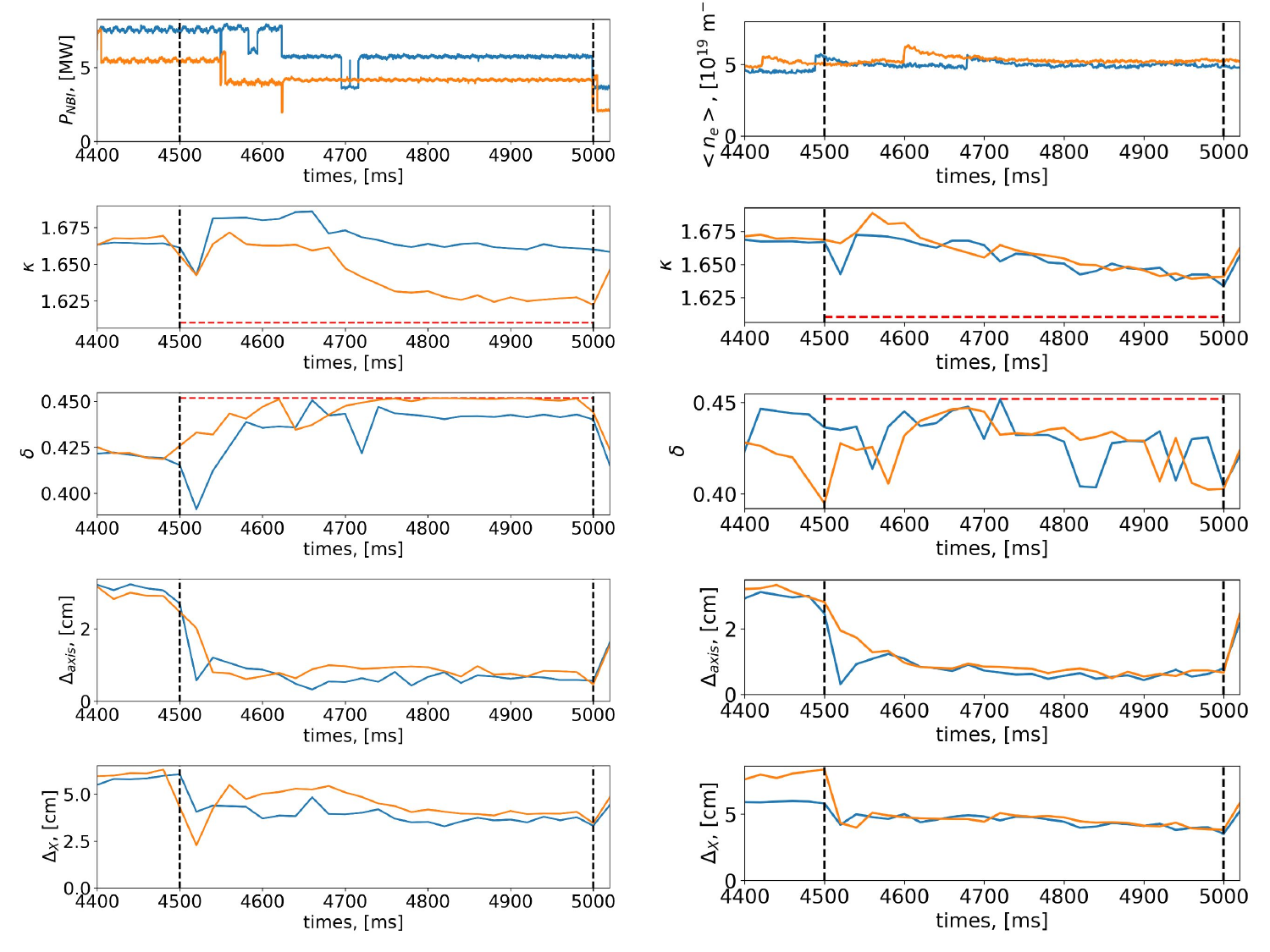}
    \caption{NBI variation (left) and pellet injection (right) effect at observed control metrics: elongation, triangularity, error in plasma center position, error in X-point position}
    \label{fig:pellets}
\end{figure}

Better control performance was demonstrated by RL25 controllers in the second testing session. Overall control error of plasma shape and center position was maintained within 1.5 cm and 1 cm, respectively, as shown in figure~\ref{fig:RL25performance}. With a different way of  $\beta_{\mathrm{N}}$ treatment, this agent demonstrated more robust behaviour in conditions of fast variations of the kinetic part of plasma equilibrium. In contrast with RL24 agents in this experiment, there is no visible effect at the plasma center position and x point position during variations of NBI power and pellet injection (see figure~\ref{fig:pellets}).

\begin{figure}
    \centering
    \includegraphics[width=0.6\linewidth]{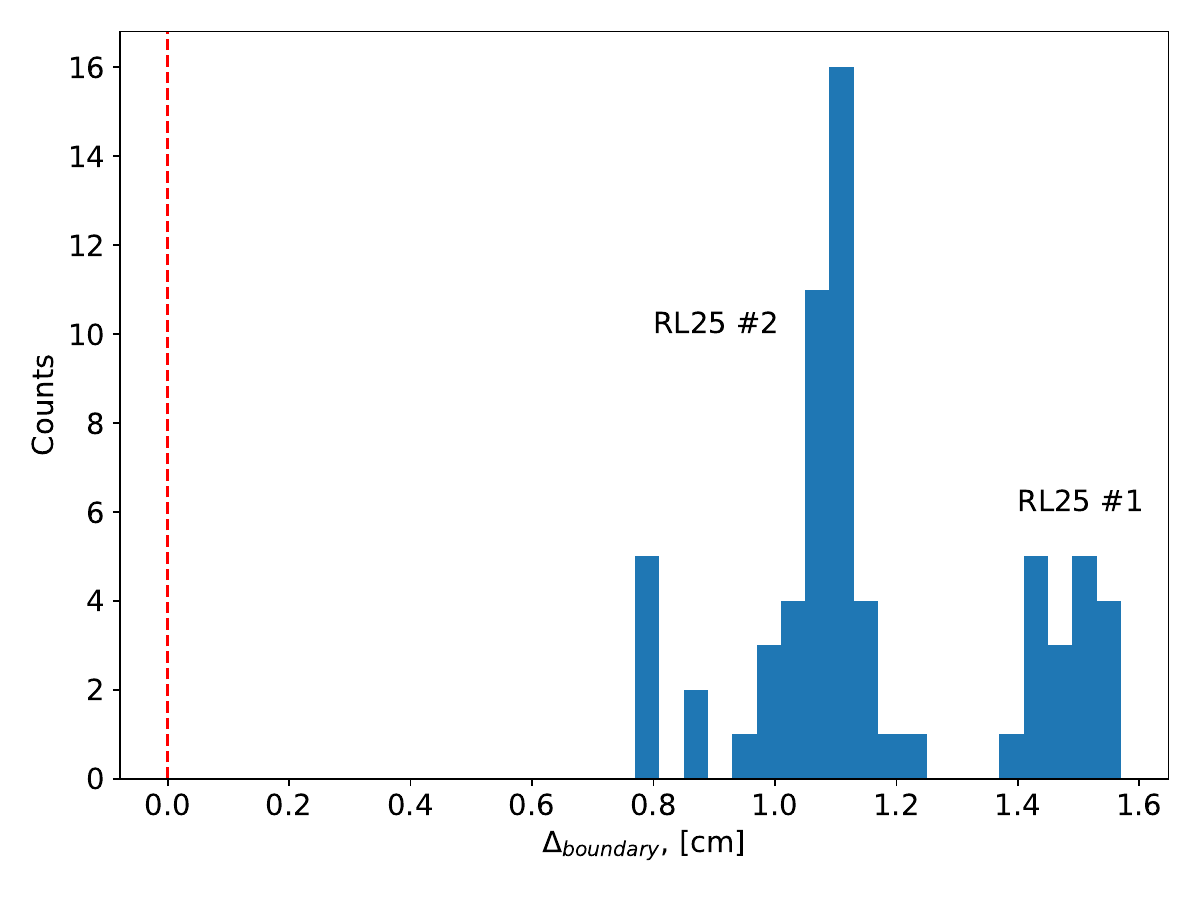}
    \caption{Control errors distribution - separatrix shape control, statistics over 11 shots with \textit{I}\textsubscript{p}=1 MA, last 100 ms to avoid transient effects influence}
    \label{fig:shapeErr}
\end{figure}

\begin{figure}
    \centering
    \includegraphics[width=0.8\linewidth]{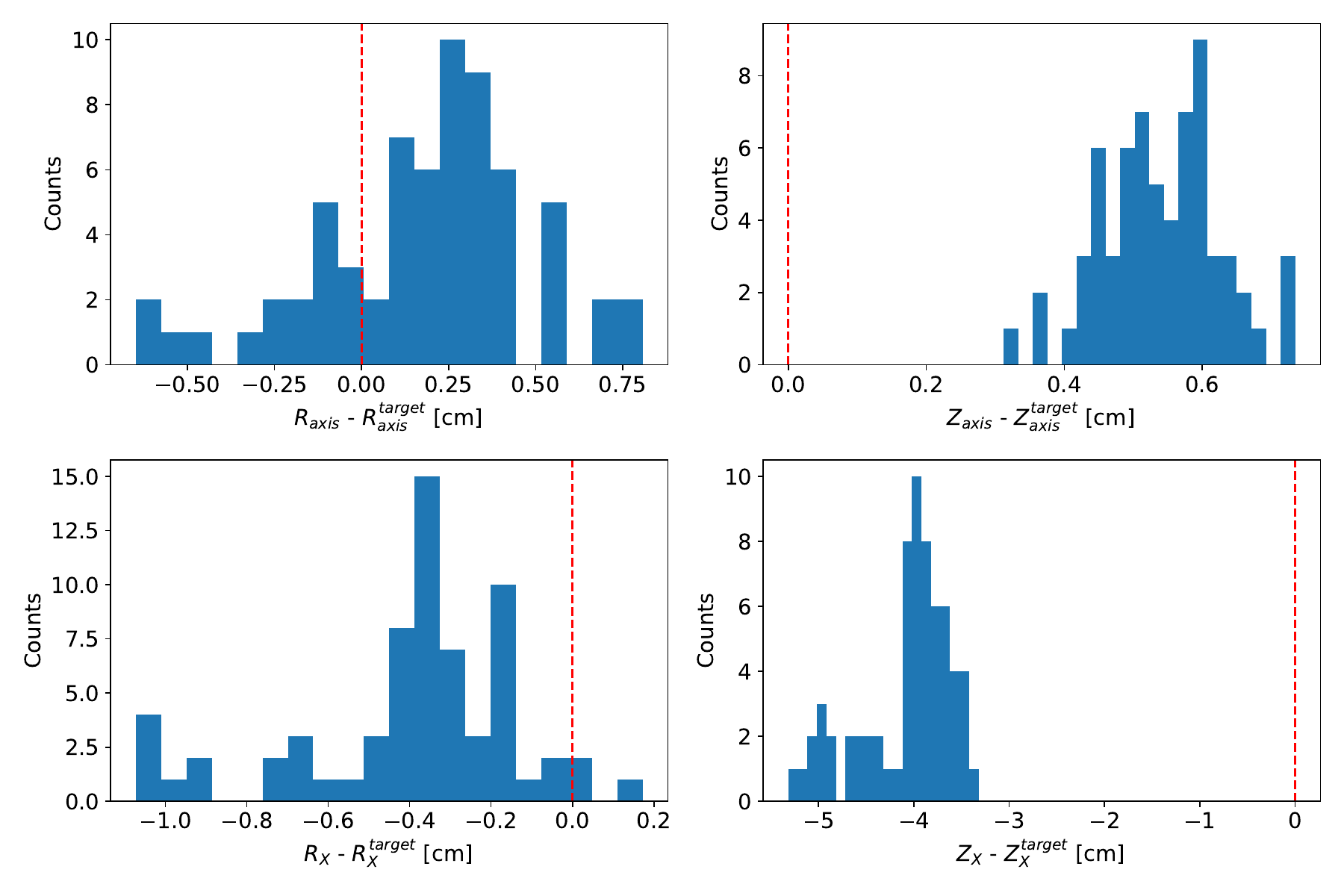}
    \caption{Control errors distribution - magnetic axis position and X-point position, statistics over 11 shots with \textit{I}\textsubscript{p}=1 MA, last 100 ms to avoid transient effects influence}
    \label{fig:posErr}
\end{figure}

Quality control can be assessed in quantitative metrics, such as error in X-point and magnetic axis horizontal and vertical position, and mean deviation of the separatrix from the target shape. These statistics were gathered using the last 100 ms of each RL25 agent activity in 11 different shots with the same target shape and X-point position. While the shape deviation metric does not provide a comprehensive understanding of the error in shape control at each point along the separatrix, it still offers a general understanding of the overall quality of RL controller performance. For most cases, the mean deviation in shape is less than 1.2 cm for RL25 \#2 as cab be seen in figure~\ref{fig:shapeErr}. Results in figure~\ref{fig:RL25performance}, demonstrate that the higher errors in shape are concentrated in the top part of the plasma column and in the vicinity of the X-point. This is further supported by the analysis of X-point positioning quality.  Statistics of errors in vertical and horizontal positioning of the plasma center and X-point by controller RL25 are presented in the figure ~\ref{fig:posErr}. Horizontal and vertical plasma center position was maintained within 0.7 cm of the reference in most cases. It is important to highlight that horizontal position control demonstrated better performance than vertical, with the error of 0.25~cm vs 0.5~cm in most cases. The same behaviour can be observed in the case of X-point, while the difference in control quality is more pronounced (0.4~cm error for horizontal position and 4~cm for vertical position). A clear maximum in statistics indicates that the achieved precision is limited mostly by the difference in modeling that was used in the training process and experimental plasma shots. Higher errors in the vertical position of the X-points are expected since the X-point location is sensitive to the balance of magnetic forces set by the plasma current density profile and the external coil currents. Deviations occur in this balance since the exact current distribution in the edge part of the plasmas was not controlled during the training procedure, affecting X-point control.

\begin{figure}
    \centering
    \includegraphics[width=0.8\linewidth]{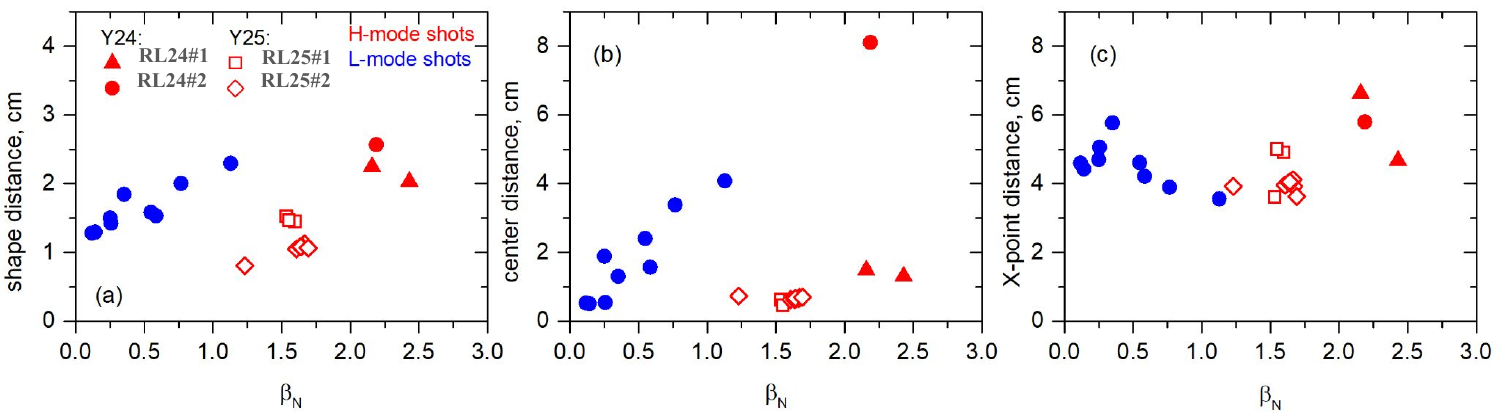}
    \caption{Change in reward components during $\beta_{\mathrm{N}}$ scan with $P_{\mathrm{NBI}}$ for both RL-controllers. While errors in shape and magnetic axis position increase with $\beta_{\mathrm{N}}$, X-point position mismatch is insensitive to it.}
    \label{fig:metrics}
\end{figure}

Reward components, such as the distances from the target for shape and coordinate of magnetic center and X-point, are given in figure~\ref{fig:metrics} as a function of the normalized ratio of the plasma pressure to the magnetic pressure $\beta_{\mathrm{N}}$, obtained using different NBI power. It can be seen in figure~\ref{fig:metrics}(a) and (b) that the mismatch in shape and magnetic center for RL24 controllers stays below 2 cm in low $\beta_{\mathrm{N}}<0.5-0.7$ (with the smallest values in Ohmic heating $P_\mathrm{NBI} = 0$~MW) and increases with $\beta_{\mathrm{N}}$. While the training procedure is designed in a way that the RL-agent effectively sees more Ohmically heated regimes, in the case of the RL24 \#1 controller, starting with H-mode high power reference helps to eliminate errors in magnetic center positions. Controllers from the RL25 series do not show the same behaviour, performing equally in all $\beta_{\mathrm{N}}$ cases where they have been tested.

\section{Discussion}\label{sec:discussion}

The new reinforcement learning-based controllers for plasma shape and position was successfully developed, integrated, and tested on the DIII-D tokamak. Experimental results demonstrate the strong applicability of the RL method for training controllers for specific device configurations using an environment that simulates both plasma response and hardware constraints. Future efforts to incorporate heating and current drive models for more accurate transport treatment will enhance the accuracy and robustness of the system, thereby creating a joint framework for scenario prediction and training advanced controllers.

RL provides powerful and efficient methods for addressing the nonlinear nature of plasma. Neural networks trained with RL can control plasma in real time, as demonstrated in the DIII-D experiments presented in this work, with the computational cost of only a few matrix operations. The flexibility of RL methods enables users to solve diverse problems from various physical domains. In recent years, magnetic control \cite{Degrave_2022, Tracey_2024, Wu_2024}, kinetic control \cite{Wakatsuki_2021, Seo_2021}, and stability control \cite{Seo_2024} were explored with RL. These promising results highlight the significant potential of RL methods for developing advanced control solutions.

RL-based control technology has the potential to address Fusion Power Plant (FPP)-relevant challenges, including the unavailability of diagnostics, limited actuators due to the blanket and shielding, and long-pulse operation. The RL approach enables exploration of the control state space through reward functions and exploratory methods, uncovering innovative solutions that might elude traditional approaches. RL approach does not require to have linear and fast model in training process, so more complex and non linear dependencies can be covered in such control methods, This fact demostrate an advantege of RL-based control in comparison with advanced Model Predictive Control methods in solving multi-controller and multi-physics control cases, From the other hand, randomization of plasma states during the training process clearly defines limits of a space of a plasma parameters, where controller is expected to be robust and perform with known quality. Considering FPP's limited operational scenario range, RL allows strictly limiting "normal operation" space, considering everything outside it as a failure mode and activating machine protection and general safety scenarios. However, advancing this technology to an FPP-compatible level requires overcoming significant challenges, particularly in the development of sophisticated plasma simulation and synthetic diagnostic codes that are both fast and accurate to meet the criteria of training time request and control accuracy at the same time.
In conclusion, it is important to note that RL, model predictive control (MPC) \cite{Maljaars_2015, Garrido_2016, Mele_2025}, and other control approaches do not exclude each other. In fact relevant application of each of these methods for different control domains will enable building a robust and efficient control framework for FPP in the future.

\section*{Acknowledgments}

This material is based upon work supported by the U.S. Department of Energy, Office of Science, Office of Fusion Energy Sciences, using the DIII-D National Fusion Facility, a DOE Office of Science user facility, under Award DE-FC02-04ER54698 and Next Step Fusion S.a.r.l. with UCSD staff supported by Next Step Fusion S.a.r.l.

\section*{Disclaimer}
This report was prepared as an account of work sponsored by an agency of the United States Government. Neither the United States Government nor any agency thereof, nor any of their employees, makes any warranty, express or implied, or assumes any legal liability or responsibility for the accuracy, completeness, or usefulness of any information, apparatus, product, or process disclosed, or represents that its use would not infringe privately owned rights. Reference herein to any specific commercial product, process, or service by trade name, trademark, manufacturer, or otherwise does not necessarily constitute or imply its endorsement, recommendation, or favoring by the United States Government or any agency thereof. The views and opinions of authors expressed herein do not necessarily state or reflect those of the United States Government or any agency thereof.

\appendix

\section{Methods}
\subsection{Simulation Environment}

A Free-boundary equilibrium and transport solver is used to model plasma response to the change of coil currents. The Magnetic equilibrium is considered to be toroidally symmetric and described by the Grad-Shafranov equation \cite{Grad_1958,Shafranov_1966}:

\begin{equation}
  R\frac{\partial}{\partial R}\left( R\frac{\partial}{\partial R} \psi \right)+ \frac{\partial ^2}{\partial Z^2}\psi = -4\pi^2 \left( \mu_0 R^2 \frac{dp}{d\psi}+F\frac{dF}{d\psi}\right)-2\pi\mu_0j_{\phi}^{\mathrm{ext}},
\label{eq:GS}
\end{equation}
where $\psi=\psi(R,Z)$ is poloidal magnetic flux, $p=p(\psi)$ is plasma pressure, $F=RB_{\phi}$ is toroidal flux function, $j_{\phi}^{\mathrm{ext}}$ is external (to the plasma) toroidal current density belonging to the currents flowing in the active coils circuits or in the passive conductors such as the vacuum vessel. The coupling of magnetic equilibrium and transport equations is performed through an iterative scheme.

Solver initialization can be done for the target condition by using reconstruction data of an old reference shot or by utilizing an inverse solver that fits coil current for a given plasma state. Transport boundary conditions (temperatures and effective ionic charge at $\rho=1$) are specified using experimental values.
\subsection{Learning Magnetic Plasma Control with Privileged Information}

We assume the agent receives control at the flat-top phase of the discharge when the plasma shape is close to the target shape. We consider the initial distribution of discharge parameters such as kinetic profiles, plasma and coil currents, plasma conductivity, electron and ion temperatures, which we denote as initial environment states $p(s_0)$. The initial states  are constructed using the experimental data from the DIII-D database. After initialization at state $s \sim p(s_0)$, the evolution of the discharge is simulated by the NSFSim. At each time step $i$ the agent receives observation $\{o_i^{rt}, o_i^{pri}\} \in \{\mathcal{O^{\mathrm{rt}}}, \mathcal{O^{\mathrm{pri}}}\}$, outputs action $a_i \in \mathcal{A}$, receives reward $r(s, a, s^{\prime}) \in \mathcal{R}$ and the environment transfers to the next state $s_{i + 1}$, where $\{\mathcal{O^{\mathrm{rt}}}\}$ is real-time observation space, and $\{\mathcal{O^{\mathrm{pri}}}\}$ is privileged observation space. We consider that the environment is partially observable since the agent only has access to a set of noisy magnetic diagnostics that can not fully describe the state of the plasma. However, we assume that the underlying environment dynamics is Markovian, i.e. the next state $s_{i + 1}$ depends only on the current environment state $s_i$ and the action of the agent $a_i$, and reward $r_i(s_i, a_i, s_{i+1})$ depends only on the current state $s_i$, action $a_i$ and next state $s_{i+1}$. 

We use the Soft Actor-Critic algorithm to train a continuous control policy. The goal of the agent is to learn a policy $\pi^{*}$ which maximizes the cumulative discounted reward plus the entropy regularization term:

\begin{equation}
    \pi^{*} = \mathrm{arg}\max_{\pi} \ex_{\tau \sim \pi} \left[ \sum_{t= 0}^{\infty} \gamma^t \left( r(s_t, a_t, s_{t+1}) + \alpha H(\pi(\cdot | o_{t}^{\mathrm{rt}}))\right)\right],
\end{equation}
where the $\gamma$ $\in$ (0, 1) is the discounting factor and entropy coefficient $\alpha > 0$ balances exploration and exploitation. The entropy of policy $\pi$ given the real-time observation $o^{\mathrm{rt}}$ is defined as follows:

\begin{equation}
    H(\pi(\cdot | o^{\mathrm{rt}})) = \ex_{a \sim \pi(\cdot | o^{\mathrm{rt}})} \left[ -\log \pi(\cdot | o^{\mathrm{rt}}) \right]
\end{equation}
The Critic network trains in parallel with the Actor network and predicts $Q$-value according to the equation below:

\begin{equation}
    Q^{\pi}(o^{pri}, a) = \ex_{\tau \sim \pi} \left[ \sum_{t = 0}^{\infty} \gamma^{t}r(s_{t}, a_t, s_{t+1}) +  \alpha \sum_{t = 1}^{\infty} \gamma^{t} H(\pi(\cdot | o_{t}^{\mathrm{rt}})) | s_0 = s, a_0 = a\right],
\end{equation} 
We assume that the privileged observation $o^{\mathrm{pri}}$ provides enough information on the state of the environment $s$, thus it can be used to predict the cumulative reward and entropy of the policy. We apply clipped-double Q-learning when training the Critic network to reduce the overestimation bias. Finally, this leads to the following learning objectives for the Actor ($\pi_{\phi}$) and Critic ($Q_{\theta}$) networks parametrized by weights $\phi$ and $\theta$:

\begin{equation}
    J_{\pi}(\phi) = \ex_{(o^{\mathrm{rt}}_t, o^{\mathrm{pri}}_t) \sim \mathcal{D}, \hat{a}_{\phi} \sim \pi(\cdot|o^{\mathrm{rt}}_t)}\left[\alpha \log \pi_{\phi}(\hat{a}_{\phi}| o^{\mathrm{rt}}_t) - \min(Q_{\theta_1}(o^{\mathrm{pri}}_t,\hat{a}_{\phi}), Q_{\theta_2}(o^{\mathrm{pri}}_t,\hat{a}_{\phi}))\right]
\end{equation}

\begin{equation}
\begin{aligned}
    J_{Q}(\theta_{1,2}) & = \ex_{(o^{\mathrm{rt}}_t, o^{\mathrm{pri}}_t, a_t, r_t, o^{\mathrm{rt}}_{t+1}, o^{\mathrm{pri}}_{t+1}) \sim \mathcal{D}, \hat{a}_{t+1} \sim \pi(\cdot|o^{\mathrm{rt}}_{t+1})}
    \left[  \frac{1}{2} \left( Q_{\theta_{1,2}}(o^{\mathrm{pri}}_t, a_t) -   \right. \right.\\
    & \left. \left. - (r + \gamma \min(Q_{\hat{\theta}_1}(o^{\mathrm{pri}}_{t+1},\hat{a}_{t+1}), Q_{\hat{\theta}_2}(o^{\mathrm{pri}}_{t+1},\hat{a}_{t+1})) - \alpha \log \pi(\hat{a}_{t+1}|o^{\mathrm{rt}}_{t+1})\right)^2  \right]
\end{aligned}
\end{equation}
where $\hat{\theta}_{1,2}$ is the smoothened copy of $\theta_{1, 2}$ weights. We use exponential moving average with smoothing parameter $\beta$ = $5 \cdot 10^{-3}$ to update $\hat{\theta}_{1,2}$ given $\theta_{1, 2}$. During training we use replay buffer $\mathcal{D}$ with capacity $10^6$, batch size 1024, Adam optimizer with learning rate $3\cdot10^{-5}$, and weight decay $0.01$, and entropy coefficient $\alpha$ = 0.2. We run 50 copies of the training environment in parallel to speed up data acquisition and perform one gradient step for each environment step. We apply the agent's control actions with time step $\Delta t$ = 1 ms, which with discount coefficient $\gamma$ = 0.9 leads to planning horizon $\Delta t / (1 - \gamma)$ = 10 ms. The maximum length of an episode during training is limited to 1000 ms.

\subsection{Rewards and Terminations}

We defined the reward function to reflect the proximity between the current and target parameters of the plasma shape. It uses three reward components $r_{\mathrm{LCFS}}$, $r_{\mathrm{mag center}}$, $r_{\mathrm{x-point}}$ defined as follows ("x" denote current value, "y" is the target value, $|\cdot|$ is euclidean distance):

\begin{equation}
\begin{aligned}
    r_{\mathrm{LCFS}} & = f\left(\frac{1}{N}\sum_{i = 1}^{N} |x_{\mathrm{LCFS}} - y_{\mathrm{LCFS}}|\right)\\
    r_{\mathrm{mag\ center}} & = f(|x_{\mathrm{mag}} - y_{\mathrm{mag}}|)\\
    r_{\mathrm{x-point}} & = f(|x_\mathrm{x-point} - y_{\mathrm{x-point}}|)
\end{aligned}
\end{equation}
Function $f$ is used to transform distance $\Delta$, such that $\Delta \leq \Delta_\mathrm{good}$ corresponds to $r = 1$:
\begin{equation}
\begin{cases}
    f(\Delta) = \mathrm{clip}(2 / (1 + \exp(-k \cdot x)), 0, 1)\\
    x = (\Delta_\mathrm{good} - \Delta) / (\Delta_\mathrm{good} - \Delta_\mathrm{bad}) 
\end{cases}
\end{equation}
We choose parameter $k = -\log(19)$, so $\Delta \geq \Delta_\mathrm{bad}$ corresponds to $r = 0.1$. We use $\Delta_\mathrm{good}$ = 0 cm and $\Delta_\mathrm{bad}$ = 8 cm in our experiments. All reward components are then aggregated using the smooth maximum function, with smoothing parameter $\alpha_{sm} = -5$, to form a scalar reward value. Hence, the agent is mostly rewarded for the smallest reward component. Additionally, to facilitate the training, we terminate episodes when the distance between the current and target plasma shape parameters is higher than 16 cm. 

\subsection{Training data}
To train RL controller for each plasma shape, we created datasets of initial plasma parameters based on DIII-D experimental data. The dataset for RL24 \#1 agent contains 119 plasma states from 5 discharges, the dataset for RL24 \#2 agent contains 1898 states from 9 discharges, the dataset for agents RL25 \#1, \#2  has 2230 states from 14 discharges. The simulation starts with the reconstructed plasma state from the dataset, after that the agent interacts with the plasma, and the evolution of the plasma is predicted by the NSFSim. On average agent needs 500k interactions with the plasma to learn a control policy. 

\subsection{Critic's effect on training performance}

\begin{figure}
    \centering
    \includegraphics[width=0.5\linewidth]{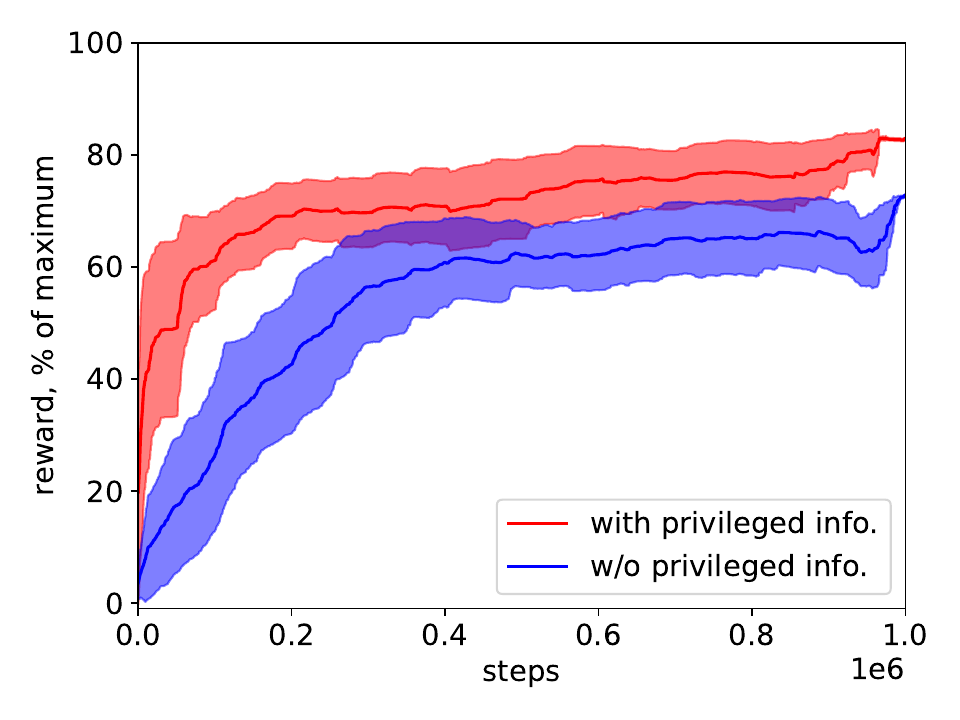}
    \caption{Cumulative reward as a function of number of training steps. RL agent trained with privileged information achieves higher reward and trains faster than without it }
    \label{fig:train}
\end{figure}

In figure~\ref{fig:train}, we demonstrate the importance of privileged information in training plasma shape controllers. It shows cumulative rewards during training averaged over 500 consecutive episodes as a function of a number of training steps. In figure~\ref{fig:train} we see that the agent trained with Critic access to privileged information achieves a higher reward and does approximately two times faster than the agent without privileged information. Both agents are trained with identical randomizations. The achieved return $\sim$70-80\% of maximum return (which equals 1 $\times$ 1000 steps) indicates strong performance of trained agents.

\clearpage

\section*{References}
\begin{thebibliography}{10}

\bibitem{Blum2008}
J~Blum, C~Boulbe, and B~Faugeras.
\newblock Real-time plasma equilibrium reconstruction in a tokamak.
\newblock {\em Journal of Physics: Conference Series}, 135:012019, November 2008.

\bibitem{Bourdelle_2015}
C.~Bourdelle, L.~Chôné, N.~Fedorczak, X.~Garbet, P.~Beyer, J.~Citrin, E.~Delabie, G.~Dif-Pradalier, G.~Fuhr, A.~Loarte, C.F. Maggi, F.~Militello, Y.~Sarazin, L.~Vermare, and JET Contributors.
\newblock L to h mode transition: parametric dependencies of the temperature threshold.
\newblock {\em Nuclear Fusion}, 55(7):073015, jun 2015.

\bibitem{nsfsim}
Randall Clark, Maxim Nurgaliev, Eduard Khairutdinov, Georgy Subbotin, Anders Welander, and Dmitri~M Orlov.
\newblock Validation of {NSFsim} as a grad-shafranov equilibrium solver at {DIII}-{D}.
\newblock {\em Fusion Eng. Des.}, 211(114765):114765, 1~February 2025.

\bibitem{Degrave_2022}
Jonas Degrave, Federico Felici, Jonas Buchli, Michael Neunert, Brendan Tracey, Francesco Carpanese, Timo Ewalds, Roland Hafner, Abbas Abdolmaleki, Diego de~las Casas, Craig Donner, Leslie Fritz, Cristian Galperti, Andrea Huber, James Keeling, Maria Tsimpoukelli, Jackie Kay, Antoine Merle, Jean-Marc Moret, Seb Noury, Federico Pesamosca, David Pfau, Olivier Sauter, Cristian Sommariva, Stefano Coda, Basil Duval, Ambrogio Fasoli, Pushmeet Kohli, Koray Kavukcuoglu, Demis Hassabis, and Martin Riedmiller.
\newblock Magnetic control of tokamak plasmas through deep reinforcement learning.
\newblock {\em Nature}, 602(7897):414–419, February 2022.

\bibitem{Eldon_2020}
D~Eldon, A~W Hyatt, B~Covele, N~Eidietis, H~Y Guo, D~A Humphreys, A~L Moser, B~Sammuli, and M~L Walker.
\newblock High precision strike point control to support experiments in the {DIII}-{D} small angle slot divertor.
\newblock {\em Fusion Eng. Des.}, 160(111797):111797, 1~November 2020.

\bibitem{Erba_1998}
M~Erba, T~Aniel, V~Basiuk, A~Becoulet, and X~Litaudon.
\newblock Validation of a new mixed bohm/gyro-bohm model for electron and ion heat transport against the {ITER}, tore supra and {START} database discharges.
\newblock {\em Nucl. Fusion}, 38(7):1013, 1~July 1998.

\bibitem{rtEFIT}
J.R Ferron, M.L Walker, L.L Lao, H.E.~St John, D.A Humphreys, and J.A Leuer.
\newblock Real time equilibrium reconstruction for tokamak discharge control.
\newblock {\em Nuclear Fusion}, 38(7):1055–1066, July 1998.

\bibitem{Garrido_2016}
Izaskun Garrido, Aitor~J. Garrido, Stefano Coda, Hoang~B. Le, and Jean~Marc Moret.
\newblock Real time hybrid model predictive control for the current profile of the tokamak à configuration variable (tcv).
\newblock {\em Energies}, 9(8), 2016.

\bibitem{Grad_1958}
H~Grad and H~Rubin.
\newblock Hydromagnetic equilibria and force-free fields.
\newblock Technical report, New York Univ., New York. Inst. of Mathematical Sciences, 10 1958.

\bibitem{SAC}
Tuomas Haarnoja, Aurick Zhou, Pieter Abbeel, and Sergey Levine.
\newblock {Soft Actor-Critic: Off-Policy Maximum Entropy Deep Reinforcement Learning with a Stochastic Actor}, 2018.

\bibitem{Holcomb_2024}
C.T. Holcomb and the DIII-D~Team.
\newblock Diii-d research to provide solutions for iter and fusion energy.
\newblock {\em Nuclear Fusion}, 64(11):112003, aug 2024.

\bibitem{Lao_2005}
L.~L. Lao, H.~E.~St. John, Q.~Peng, J.~R. Ferron, E.~J. Strait, T.~S. Taylor, W.~H. Meyer, C.~Zhang, and K.~I. You.
\newblock Mhd equilibrium reconstruction in the diii-d tokamak.
\newblock {\em Fusion Science and Technology}, 48(2):968--977, 2005.

\bibitem{Maljaars_2015}
E.~Maljaars, F.~Felici, M.R. de~Baar, J.~van Dongen, G.M.D. Hogeweij, P.J.M. Geelen, and M.~Steinbuch.
\newblock Control of the tokamak safety factor profile with time-varying constraints using mpc.
\newblock {\em Nuclear Fusion}, 55(2):023001, jan 2015.

\bibitem{Mele_2025}
Adriano Mele, Alessandro Tenaglia, Federico Felici, Cristian Galperti, Daniele Carnevale, Stefano Coda, Antoine Merle, Alfredo Pironti, and Olivier Sauter.
\newblock Design and implementation of a model-based hierarchical architecture for plasma shape control in the tcv tokamak.
\newblock {\em Plasma Physics and Controlled Fusion}, 2025.

\bibitem{Moret2015}
J.-M. Moret, B.P. Duval, H.B. Le, S.~Coda, F.~Felici, and H.~Reimerdes.
\newblock Tokamak equilibrium reconstruction code liuqe and its real time implementation.
\newblock {\em Fusion Engineering and Design}, 91:1–15, February 2015.

\bibitem{Seo_2024}
Jaemin Seo, Sangkyeun Kim, Azarakhsh Jalalvand, Rory Conlin, Andrew Rothstein, Joseph Abbate, Keith Erickson, Josiah Wai, Ricardo Shousha, and Egemen Kolemen.
\newblock Avoiding fusion plasma tearing instability with deep reinforcement learning.
\newblock {\em Nature}, 626(8000):746--751, 21~February 2024.

\bibitem{Seo_2021}
Jaemin Seo, Y-S Na, B~Kim, C~Y Lee, M~S Park, S~J Park, and Y~H Lee.
\newblock Feedforward beta control in the {KSTAR} tokamak by deep reinforcement learning.
\newblock {\em Nucl. Fusion}, 61(10):106010, 2~September 2021.

\bibitem{Shafranov_1966}
V.D. Shafranov.
\newblock Plasma equilibrium in a magnetic field.
\newblock {\em Reviews of Plasma Physics}, 2:103–151, 1966.

\bibitem{Thome_2024}
K~E Thome, M~E Austin, A~Hyatt, A~Marinoni, A~O Nelson, C~Paz-Soldan, F~Scotti, W~Boyes, L~Casali, C~Chrystal, S~Ding, X~D Du, D~Eldon, D~Ernst, R~Hong, G~R McKee, S~Mordijck, O~Sauter, L~Schmitz, J~L Barr, M~G Burke, S~Coda, T~B Cote, M~E Fenstermacher, A~Garofalo, F~O Khabanov, G~J Kramer, C~J Lasnier, N~C Logan, P~Lunia, A~G McLean, M~Okabayashi, D~Shiraki, S~Stewart, Y~Takemura, D~D Truong, T~Osborne, M~A Van~Zeeland, B~S Victor, H~Q Wang, J~G Watkins, W~P Wehner, A~S Welander, T~M Wilks, J~Yang, G~Yu, L~Zeng, and the DIII-D~Team.
\newblock Overview of results from the 2023 DIII-D negative triangularity campaign.
\newblock {\em Plasma Physics and Controlled Fusion}, 66(10):105018, sep 2024.

\bibitem{Tracey_2024}
Brendan~D Tracey, Andrea Michi, Yuri Chervonyi, Ian Davies, Cosmin Paduraru, Nevena Lazic, Federico Felici, Timo Ewalds, Craig Donner, Cristian Galperti, Jonas Buchli, Michael Neunert, Andrea Huber, Jonathan Evens, Paula Kurylowicz, Daniel~J Mankowitz, and Martin Riedmiller.
\newblock Towards practical reinforcement learning for tokamak magnetic control.
\newblock {\em Fusion Eng. Des.}, 200(114161):114161, 1~March 2024.

\bibitem{Wagner_2007}
F~Wagner.
\newblock A quarter-century of h-mode studies.
\newblock {\em Plasma Physics and Controlled Fusion}, 49(12B):B1, nov 2007.

\bibitem{Wakatsuki_2021}
T~Wakatsuki, T~Suzuki, N~Oyama, and N~Hayashi.
\newblock Ion temperature gradient control using reinforcement learning technique.
\newblock {\em Nucl. Fusion}, 61(4):046036, 1~April 2021.

\bibitem{Walker_2003}
M.L. Walker, J.R. Ferron, D.A. Humphreys, R.D. Johnson, J.A. Leuer, B.G. Penaflor, D.A. Piglowski, M.~Ariola, A.~Pironti, and E.~Schuster.
\newblock Next-generation plasma control in the diii-d tokamak.
\newblock {\em Fusion Engineering and Design}, 66-68:749--753, 2003.
\newblock 22nd Symposium on Fusion Technology.

\bibitem{Welander_2019}
Anders Welander, Erik Olofsson, Brian Sammuli, Michael~L. Walker, and Bingjia Xiao.
\newblock Closed-loop simulation with grad-shafranov equilibrium evolution for plasma control system development.
\newblock {\em Fusion Engineering and Design}, 146:2361--2365, 2019.
\newblock SI:SOFT-30.

\bibitem{Wu_2024}
Niannian Wu, Zongyu Yang, Rongpeng Li, Ning Wei, Yihang Chen, Qianyun Dong, Jiyuan Li, Guohui Zheng, Xinwen Gong, Feng Gao, Bo~Li, Min Xu, Zhifeng Zhao, and Wulyu Zhong.
\newblock {High-Fidelity Data-Driven Dynamics Model for Reinforcement Learning-based Magnetic Control in HL-3 Tokamak}, 2024.

\end{thebibliography}
\end{document}